\documentclass[]{spie}  
\usepackage[]{graphicx}
\usepackage{color} 
\usepackage{floatflt}
\usepackage{multirow}
\usepackage{slashbox}

\def\lesssim{\mathrel{\hbox{\rlap{\hbox{\lower4pt\hbox{$\sim$}}}\hbox{$<$}}}}
\def\gtsim{\mathrel{\hbox{\rlap{\hbox{\lower4pt\hbox{$\sim$}}}\hbox{$>$}}}}

\def\farcs{\hbox{$.\!\!^{\prime\prime}$}}

\newcommand{\ci}{{\sc C i}}
\newcommand{\hi}{{\sc H i}}
\newcommand{\pks}{PKS 1830$-$211}

\newcommand{\jyb}          		{\ensuremath{{\rm Jy~beam^{-1}}}}

\title{The eSMA: description and first results} 

\author{Sandrine Bottinelli\supit{a},
Ken H. Young\supit{b},
Richard Chamberlin\supit{c},
Remo P.J. Tilanus\supit{d,i},
Mark A. Gurwell\supit{b},
Dave J. Wilner\supit{b},
Hiroko Shinnaga\supit{c},
Hiroshige Yoshida\supit{c},
Per Friberg\supit{d},
Huib Jan van Langevelde\supit{e,a},
Ewine F. van Dishoeck\supit{a,f},
Michiel R. Hogerheijde\supit{a},
A. Meredith Hughes\supit{b},
Robert D. Christensen\supit{h},
Richard E. Hills\supit{g},
John S. Richer\supit{g},
Emily Curtis\supit{g},
and the eSMA commissioning team
\skiplinehalf
\supit{a} Leiden Observatory, Leiden University, P.O. Box 9513, NL-2300 RA Leiden, The Netherlands; 
\skiplinehalf
\supit{b} Harvard-Smithsonian Center for Astrophysics, MS 42, 60 Garden Street, Cambridge, MA 02138, USA; 
\skiplinehalf
\supit{c} Caltech Submillimeter Observatory Office, 111 Nowelo St., Hilo HI 96720, USA;
\skiplinehalf
\supit{d} Joint Astronomy Center, 660 N. A'ohoku Place, University Park, Hilo HI 96720, USA; 
\skiplinehalf
\supit{e} Joint Institute for VLBI in Europe, Radiosterrenwacht Dwingeloo, Postbus 2, NL-7990 AA Dwingeloo, The Netherlands;
\skiplinehalf
\supit{f} Max-Planck-Institut f\"{u}r Extraterrestrische Physik, Postfach 1312, 85741 Garching, Germany; 
\skiplinehalf
\supit{g} Cavendish Astrophysics Group, Cavendish Laboratory, University of Cambridge,
JJ Thomson Avenue, Cambridge CB3 0HE, UK; 
\skiplinehalf
\supit{h} Harvard-Smithsonian Center for Astrophysics, Submillimeter Array, 645 North AÕohoku Place, Hilo HI 96721, USA; 
\skiplinehalf
\supit{i} Netherlands Organisation for Scientific Research, Laan van Nieuw Oost-Indie 300, NL-2509 AC The Hague, The Netherlands
}

\authorinfo{Further author information: (Send correspondence to S.B.)\\ %
S.B.: E-mail: sandrine@strw.leidenuniv.nl, Telephone: +31 (0)71 527 8433
}

\begin{document} 
  \maketitle 

\begin{abstract}

The eSMA (``extended SMA'') combines the SMA, JCMT and CSO into a
single facility, providing enhanced sensitivity and spatial resolution
owing to the increased collecting area at the longest
baselines. 
Until ALMA early science observing (2011), 
the eSMA will be the facility capable of
the highest angular resolution observations at 345~GHz. The gain in
sensitivity and resolution will bring new insights in a variety of
fields, such as protoplanetary/transition disks, high-mass star
formation, solar system bodies, nearby and high-$z$ galaxies. Therefore
the eSMA is an important facility to prepare the grounds for ALMA and
train scientists in the techniques.

Over the last two years, and especially since November 2006, there has
been substantial progress toward making the eSMA into a working
interferometer. In particular, (i) new 345-GHz receivers, that match
the capabilities of the SMA system, were
installed at the JCMT and CSO; (ii) numerous tests
have been performed for receiver, correlator and baseline calibrations
in order to determine and take into account the effects arising from
the differences between the three types of antennas; (iii) first
fringes at 345~GHz were obtained on August 30 2007, and the array has
entered the science-verification stage.

We report on the characteristics of the eSMA and its measured
performance at 230~GHz and that expected at 345 GHz. We also present
the results of the commissioning and some initial science-verification
observations, including the first absorption measurement of the C/CO
ratio in a galaxy at $z$=0.89, located along the line of sight to the
lensed quasar \pks, and on the imaging of the vibrationally excited
HCN line towards IRC+10216.

\end{abstract}




\section{INTRODUCTION}
\label{sec:intro}  

The submillimeter window is critically important for astronomy because
it probes cold and dense environments whose spectral energy
distribution peaks in this wavelength range.  Moreover, the
submillimeter window offers unique access to numerous high-excitation
lines of molecules. With its dry atmosphere, Mauna Kea is a privileged
location to open this window and peer into the dense, dust-enshrouded
regions of the Universe.  The Caltech Submillimeter Observatory (CSO)
and James Clerk Maxwell Telescope (JCMT) were among the first
submillimeter facilities worldwide and have allowed astronomers to
obtain novel results with undiminished success since their first-light
on Mauna Kea in the mid-eighties.

Despite their numerous advantages, these single-dishes have a limited
angular resolution and the
SubMillimeter Array (SMA), the first imaging
interferometer at submillimeter wavelengths,
 was built on Mauna Kea with the goal
of providing high-angular resolution observations in this regime.
The SMA consists of
8$\times$6-m dishes provided by the Smithsonian
Astrophysical Observatory (SAO, USA) and ASIAA (Taiwan). 
Following the successful realization of using the 10.4-m
CSO and 15-m JCMT as a two-antenna interferometer 
(1992-1995, see Ref.~\citenum{lay-etal94,lay-etal95}), it was foreseen,
as the first SMA antennas were brought to Mauna Kea almost a decade ago,
to combine the SMA with the JCMT and the CSO into an ``extended SMA'' or
eSMA, providing enhanced sensitivity and spatial resolution
compared with the SMA. The JCMT-SMA Memorandum of Understanding was
signed in 2001; the CSO later agreed to join the project. Over the
last two years, there has been substantial progress toward making the
eSMA into a working interferometer.

The eSMA is now in a position to start fully exploiting the advantages which
motivated the three individual observatories to embark on this joint venture:
\begin{itemize}
\item {\it Increased resolution:} With the CSO in the array, the longest baseline extends to
782~m, compared to 509~m for the SMA in very-extended configuration. This corresponds
to an angular resolution better than $\sim0\farcs2$ at 345~GHz.
\item {\it Increased sensitivity:} The addition of the CSO and JCMT doubles the collecting
area of the SMA alone from 226 to 488~m$^2$. 
Together with dual polarization  
(and depending on the assumptions made for the values of the system temperatures in
 Eq.~\ref{eq:sigma} of Appendix~\ref{ap:sensitivity}), this
 provides an improvement of up to a factor of $\sim3-5$ in observing speed, compared to the
 current, single polarized SMA. 
 Moreover, 
 this extra collecting area is located on the longest baselines /
highest angular resolution, where it is most critically needed.
\end{itemize}

\begin{figure}
   \begin{center}
\begin{minipage}[c]{0.62\linewidth}
\centering
   \includegraphics[width=\linewidth]{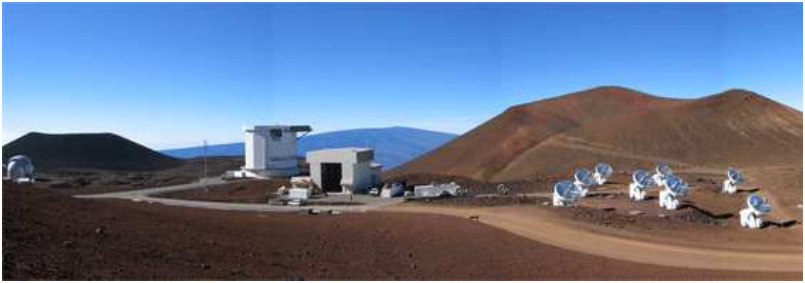} 
\end{minipage}%
\hfill%
\begin{minipage}[c]{0.36\linewidth}
\centering
   \includegraphics[width=\linewidth]{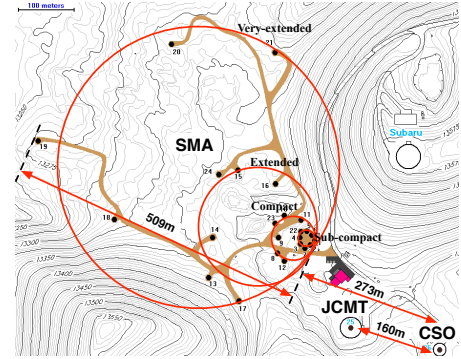}    
\end{minipage}
   \end{center}
   \caption[] 
   { \label{mk} 
   {\it Left:} From left to right, the CSO, JCMT, SMA control building and SMA antennas. --- 
   {\it Right:} Topographic map showing the pad positions
   for the SMA antennas, and the locations of the JCMT and CSO. The red circles are indicative
   of the four main different SMA configurations. Also labeled are a few key distances.   }
\end{figure}

In this paper, we describe the technical challenges encountered in the
process of joining the CSO and JCMT to the SMA, outline some sciences
objectives, and present the first results obtained with this facility.


\section{TECHNICAL DESCRIPTION} 

The CSO, JCMT and SMA, which together make the eSMA, are pictured in the left panel of Fig.~\ref{mk},
while the right panel shows the locations of the JCMT and CSO and the pad-positions for the SMA antennas.
In this section, we describe the hardware and operational issues inherent to the eSMA, as well as 
some of the tests that have been carried out to try and solve the problems intrinsic to this
heterogeneous array. Note that although the eSMA is designed to
operate in the 320-355~GHz window, the testing to date has been done at
lower frequencies (230-270~GHz). 
   
\subsection{Hardware}

\subsubsection{Adding the JCMT and CSO to the SMA array} 

\begin{figure}[ht]
   \begin{center}
   \includegraphics[width=0.37\linewidth]{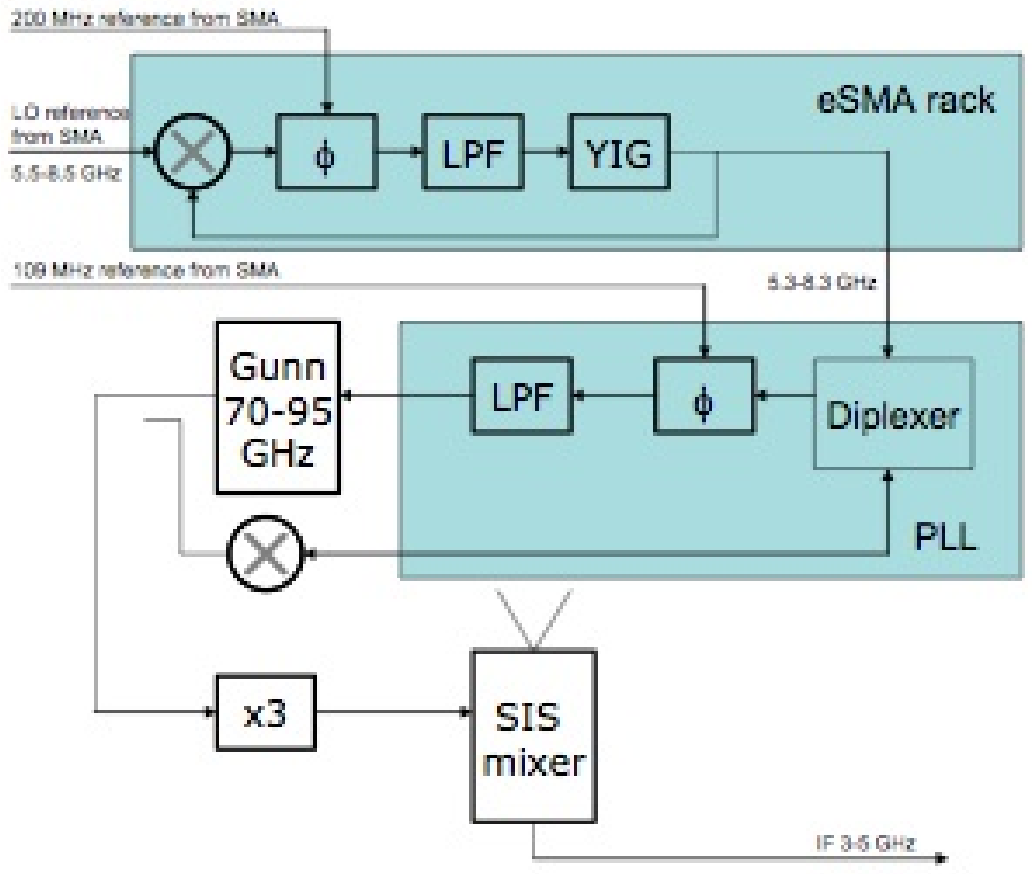} \hfill   \includegraphics[width=0.57\linewidth]{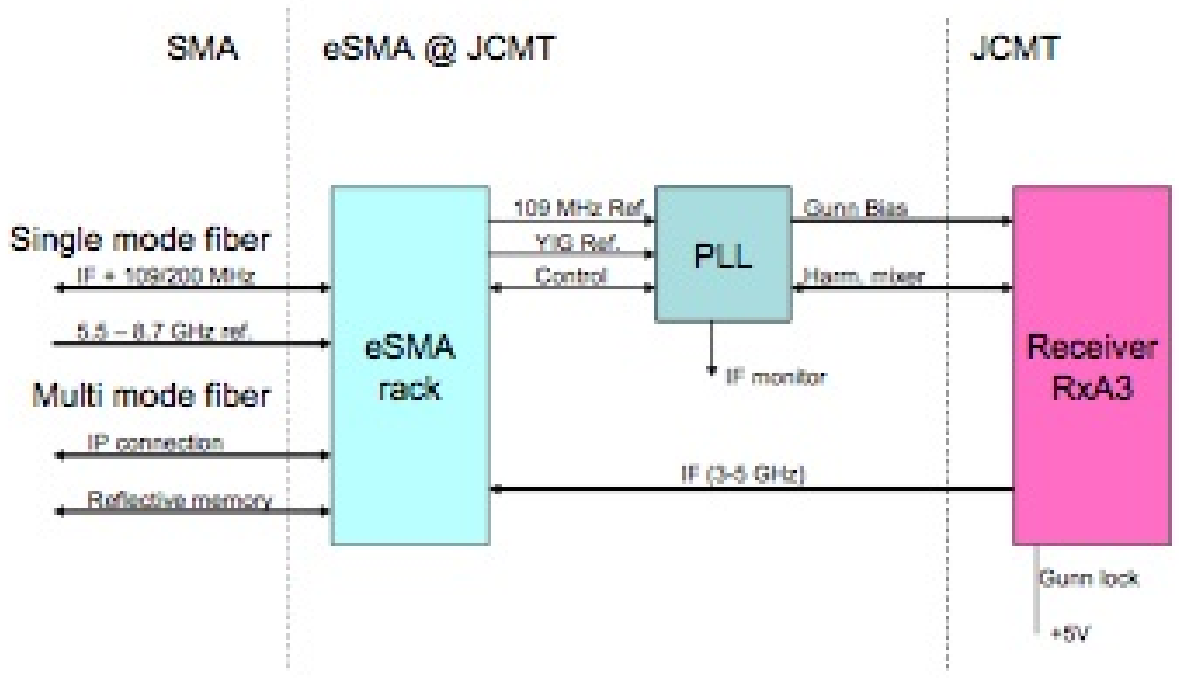} 
   \end{center}
   \caption[] 
   { \label{connections} 
   {\it Left:} Schematic view of the LO and PLL circuit. --- 
   {\it Right:} Overview of the eSMA crate connections between the JCMT and SMA.  
   Similar connections are present between the CSO and SMA. }
\end{figure}

Sumitomo single mode fibers were pulled from the SMA control building
and spliced into the existing fiber between the CSO and JCMT in
2000. These fibers distribute the LO reference signals from the SMA to
the CSO and JCMT receivers and carry the returning IF signal back to the
SMA correlator.
In addition, multi-mode fiber bundles were installed
for computer communications and control. 
Following the commissioning of the SMA itself, 
SMA IF/LO conditioning and control electronics were
duplicated and installed at the CSO and the JCMT. 
The above steps were absolutely necessary in order to have 
identical LO chains for all the telescopes involved, which
is essential for eSMA operations, as a fixed cable length is assumed
in the correlation.\\
Furthermore,
the fibers and
attached equipment have been configured to reside within the SMA
network, transparently extending the array to include the CSO and
JCMT antennas. Schematic views of the LO/PLL circuit and of the
crate connections are shown in Fig.~\ref{connections}.

SMA control to the JCMT and the CSO is channeled  
through dedicated computers at each telescope.  The dedicated  
computers also control the SMA specific IF/LO conditioning electronics.
They receive
commands through Remote Procedure Call (RPC)
and reflective memory, which is also used to
report back the antenna status.
With only minor differences, control of the SMA Antenna equipment at
the JCMT and CSO is identical to the rest of the array. However, the
receivers, telescope, and control software of the JCMT and CSO are
different from those at the SMA and each other and require a custom
interface. A simplified ASCII command set, implementable via e.g. a
serial line connection, was designed as the common interface between
the observatories. This effectively allows the SMA to take remote
control of the JCMT and CSO and monitor status information such as tracking errors.
The latter is important because the different
telescopes have different slew-rates and cable-wraps, resulting in a
different routing and arrival time on target; individual baselines
are flagged until both telescopes are on target. Using the interface
the SMA can also control e.g. the switching of loads in front of the
receivers in order to carry out calibrations. The implementation of
the full interface command-set is ongoing as of summer 2008.

The eSMA required a modification of the correlator software at
the SMA. As the SMA is designed to operate in dual-frequency mode with
8 antennas, there is enough hardware for running the eSMA 10-antenna
single-frequency configuration (45 baselines). 
Since August 2007, the eSMA has been observing successfully
with all 45 baselines.


\subsubsection{Heterogeneous array issues}
\label{issues}

For all practical purposes, the SMA antennas are identical, so that
telescope-specific issues typically cancel out. The CSO and
JCMT, however, are of a completely different design resulting in a
more complicated situation requiring special corrective measures.

Heterodyne receivers used in submillimeter astronomy observe a single
linear polarization and, because of the different optical
configuration, each type of telescope observes a different
polarization while pointed to the same target. To correct for this
half-wave plates are installed in the beam at the CSO and JCMT that
automatically rotate to a common polarization as a function of elevation.

The rotation of the plates introduces elevation dependent phase
terms. Additional elevation dependent phase terms arise from purely
mechanical issues.  Examples are a pronounced non-intersection of axes
term at the CSO and a measureable drop of the elevation axis at the
JCMT when moving from horizon to zenith. Unfortunately such phase
terms are degenerate with regular baseline terms and their separation
is time-consuming, requiring dedicated and repeated tracking of sources
differing e.g. only in elevation (see section \ref{bl tests}). Presently the eSMA is converging
towards the required maximum residual baseline uncertainty of $\sim$0.1~mm.

The observing software at the SMA also required modifications to take
the differences between the telescopes into regard: e.g. for
interferometric pointing observations the smaller beams of the CSO and
JCMT need to be accounted for.

\subsection{eSMA operations}

Operations of the eSMA are completely controlled
from the SMA side. The SMA is typically staffed during the first half
of the night and during the second
half, it is controlled completely remotely either from
Cambridge, Massachusetts  (4 out of 5 weekdays) or from 
Taipei, Taiwan (1 out of 5). 
However, neither the CSO nor the JCMT are set up for the level
of remote, un-attended observing as is usual for the SMA. Currently
during eSMA commissioning both the CSO and JCMT are independently
staffed. Once the commissioning has finished it is expected to support eSMA
observations with a single shared team for the CSO and JCMT.  The
requirement for such a team to remain available at the summit
throughout the night will likely be dictated more by security and
safety issues than technical constraints.

\subsection{Baselines effects}
\label{bl tests}

As mentioned in section \ref{issues}, a significant amount of time has
been spent during the testing phase to perform
necessary array calibration observations. These observations are
   designed to determine accurately the relative positions of the
   individual elements (``baselines'') along with antenna-dependent
   terms (non-intersection of the axes, etc.).  All these terms affect
   in particular the ability to transfer phase calibration solutions
   from calibrators to target sources and therefore, ultimately, our
   ability to accurately image a region in the sky. To prevent this potential issue, 
   these baseline- and antenna-dependent terms
   must be determined to a fraction of a wavelength (typically
   0.1~mm). 

\begin{figure}[h]
   \begin{center}
   \begin{tabular}{cc}
   \includegraphics[width=0.45\linewidth]{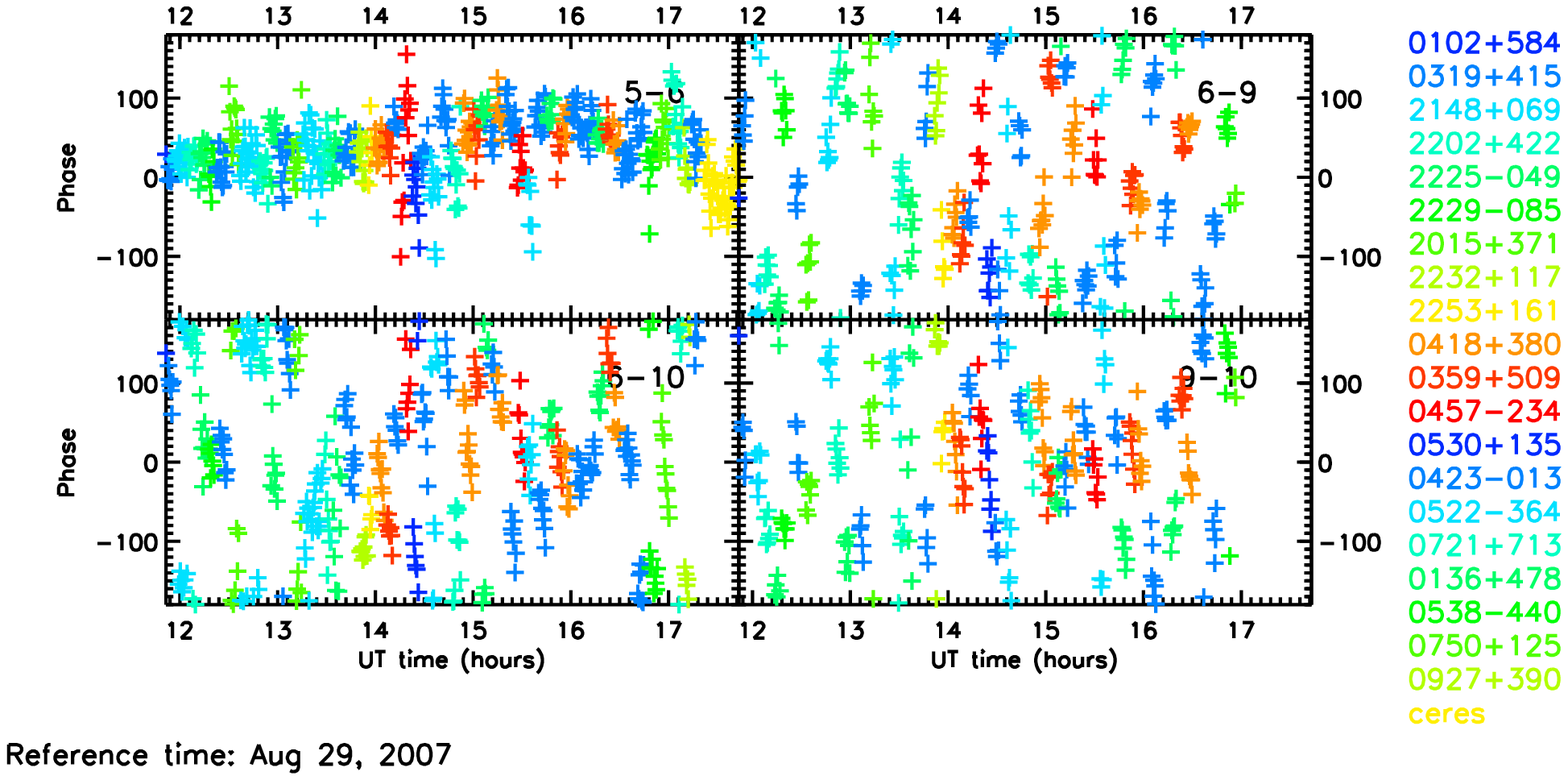} &
   \includegraphics[width=0.45\linewidth]{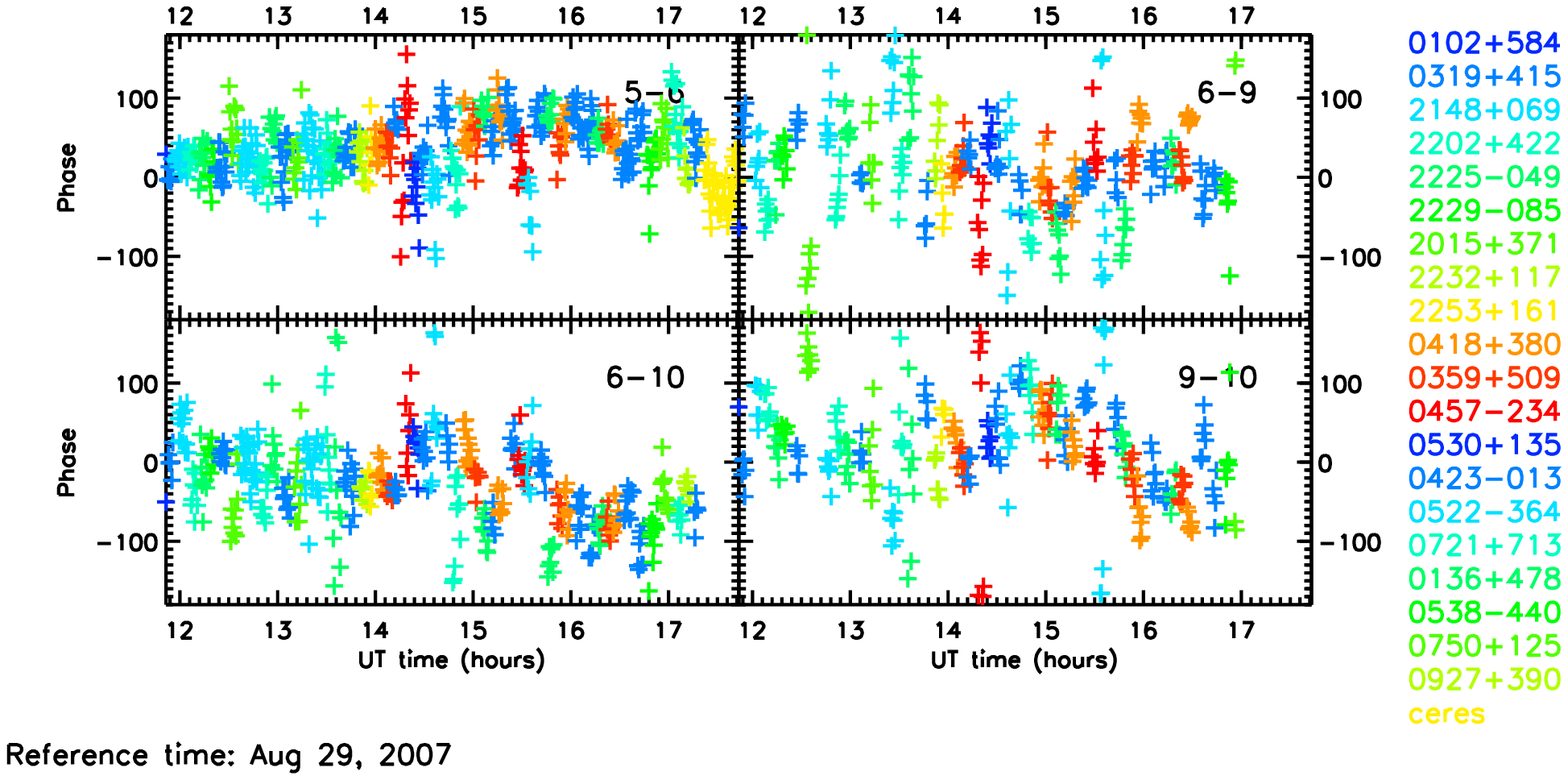}
   \end{tabular}
   \end{center}
   \caption[] 
   { \label{bl sol} 
   {\it Left:} Observed amplitude and phase for SMA-SMA (5-6), SMA-JCMT (6-10), SMA-CSO (6-9)
   and CSO-JCMT (9-10) baselines during the first baseline track obtained on 2007 August 29. ---
   {\it Right:} Effect of applying the solution derived from fitting the baseline equation 
   (see Appendix~\ref{bl eq reg}) to the baseline-track data. }
\end{figure}

For an homogeneous array, the relative antenna positions are typically obtained from observations
of unresolved calibration sources for which the positions are known. The phase
of such calibrators should ideally be zero but due to the uncertainties in the positions,
it is in fact given by Eq.~\ref{eq reg} of Appendix~\ref{bl eq reg}. 
Observations of sources spanning a large range of hour angles and
declinations hence allow us to derive the uncertainties in the $X$, $Y$ and $Z$ respectively.
Figure~\ref{bl sol} shows the effect of solving for the relative positions.
The initial values found from these August 29, 2007 observations were 
($\Delta X$, $\Delta Y$, $\Delta Z$) = 
10.25, $-$8.53, 14.11~mm
and $-$8.82, 3.14, $-$22.96~mm
for the JCMT and CSO, respectively. 

However, for an heterogeneous array, these solutions are degenerate as
they encompass antenna-dependent terms as well as the relative
positions.  Therefore, additional tests have been performed, for which
a carefully chosen subset of quasars have been observed.  For example,
a pair of quasars with similar right ascension but opposite
declination can allow determination of elevation effects (e.g. those
due to the non-intersection of the azimuth and elevation axes -- see
Appendix~\ref{bl eq el}). Indeed, in this case, the difference between
the phases of two such quasars is independent of the uncertainties in
the $X$ and $Y$ positions and can be fitted with elevation terms
only.Another appropriate choice is a pair with similar declination
and different R.A. In this case, using the functional form of the
dependence of the phase on the elevation derived with the first pair
of quasars, this second pair then provides a good solution for the $X$
and $Y$ positions.

\subsection{eSMA properties and (expected) performances}

Table~\ref{specs} summarizes the capabilities of the eSMA
in comparison with the SMA alone, both in single (sp) and dual (dp) polarization modes.
Estimates for the sp mode were obtained from the SMA sensitivity calculator:
{\sf http://sma1.sma.hawaii.edu/beamcalc.html}, 
assuming a frequency of 345~GHz, 6.7~hr of on-source integration time (H.A.=$\pm$5), 
a source declination of 25$^\circ$, and good atmospheric conditions (pvw=1.5~mm). 
Factors between sp and dp modes can be found in Appendix~\ref{ap:sensitivity}.

\begin{table}[h]
\begin{center}
\caption{Specifications and sensitivities of the eSMA and SMA arrays.}
\label{specs}
\begin{tabular}{|l|c|c|c||c|c|}
\hline
Array & eSMA$_{\rm sp}$ & eSMA$_{\rm sp}$ & eSMA$_{\rm dp}$ & SMA$_{\rm sp}$ & SMA$_{\rm dp}$  \\ 
(Band in GHz) & (230) & (345) & (345) & (230) & (345) \\
\hline
Frequency range (GHz) & 210--280 & 315--380 & 320--355 & 272--349 & 320--355 \\
\hline
Number of baselines & 45 & 45 & 28 & 28 & 28 \\
\hline
Total collecting area (m$^2$) & 488 & 488 & 431 & 226 & 226 \\
\hline
Longest baseline (m) & \multicolumn{3}{c||}{ 782 } & \multicolumn{2}{c|}{ 509 }  \\
\hline
Resolution (arcsec) & {\bf 0.22}$\times${\bf 0.55} & \multicolumn{2}{c||}{ 0.18$\times$0.32 } 
& \multicolumn{2}{c|}{ 0.24$\times$0.32 } \\
\hline
Resolution (AU at 150 pc) & $\sim${\bf 30-80} & \multicolumn{2}{c||}{ $\sim$25-50 } 
& \multicolumn{2}{c|}{  $\sim$35-50 }  \\
\hline
       Continuum (mJy/beam) & {\bf 2.1} & 1.1 & 0.9 & 2.3 & 1.6 \\ 
\hline
Line (K per 1 km/s channel) & {\bf 12.4} & 5.6 & 4.6 & 10.7 & 7.6 \\ 
\hline
\end{tabular}
\end{center}
{\sc notes.} --- {\it (i)} sp and dp denote single and dual polarization modes respectively ;
eSMA$_{\rm sp}$ is 8$\times$6~m + 10.4~m + 15~m while 
eSMA$_{\rm dp}$ is 6$\times$6~m + 10.4~m + 15~m 
(see text for details). 
{\it (ii)} Resolutions are given for uniform weighting and sensitivities for natural weighting.
--- {\it (iii)} Bold characters denote 
values measured for the 7-hr track on \pks\ (see Sec.~\ref{pks}); note that those
data were obtained under worse (pvw$\sim$3-4~mm) weather conditions than assumed 
for the estimates, and were reduced using uniform weighting.
\end{table}

Note that the eSMA currently operates in single polarization mode.  
The  JCMT has the capability to combine with the SMA in dual  
polarization, once the SMA will be fully equipped with the appropriate receivers.   
Within one or two years the CSO may also have that  capability.
In the event of dual-polarization observations with the eSMA, two SMA antennas
will have to be dropped from the array due to correlator capacities.

\section{SCIENCE OBJECTIVES} 

The eSMA will address a number of key issues in a large range
astrophysics topics, from Solar
System objects and galactic low- and high-mass star-forming regions to
high-$z$ galaxies.  Science projects that will benefit the most from
eSMA compared with the SMA alone are those that fall into the
following categories: (i) detecting and imaging faint continuum sources
(of order a few mJy) at subarcsec resolution; (ii) detecting and
imaging thermal ($\gtsim$ 30~K) lines tracing dense gas at subarcsec
resolution; (iii) providing larger samples of moderately bright
sources for statistical analysis; and (iv) providing the best possible
data on key template sources prior to ALMA as a reference.
 
Examples of science cases where the eSMA will have an impact have been
described in a report of a workshop held at Leiden Observatory, the Netherlands
in early 2007. Information can be found at:\\
{\sf http://www.leidenuniv.nl/$\sim$bottinelli/esma-workshop/report\_esma\_workshop.pdf}.\\
These science cases include the following:

\begin{itemize}
\item{\it Solar System objects.} --- The high sensitivity and angular
  resolution of eSMA will enable important studies of Solar System
  objects to be carried out, such as comets, Kuiper Belt Objects (KBOs),
  trans-Neptunian objects and Centaurs.  These objects are
  likely to consist of the most unaltered matter from the Solar
  Nebula (see Ref.~\citenum{bockelee-morvan-etal01}, and references therein),
  and are therefore essential to our understanding of the young Solar System.
  In particular, areas in which eSMA observations will provide crucial information
  are: (i) jet features from the nuclei of comets; 
  (ii) atmospheric composition of, e.g., Pluto, Triton, KBOs;
  (iii) spatial variability in Io's atmosphere and the potential link with volcanic mechanisms;
  (iv) kinematics of atmospheres (presence of ``winds'', e.g. in Titan);
  (v) light-curves of asteroids.

\item{\it Protoplanetary disks.}--- Circumstellar disks are an
  ubiquitous feature occurring in the star formation process, due to the
  conservation of angular momentum as the ISM
condenses from cloud
  core to protostellar scales. Planet formation begins in the gas-rich
  disks
around young pre-main sequence stars.  The density,
  temperature and velocity distribution of these disks critically
  determines the growth of dust particles, affects the chemistry and
  coagulation
of planetesimals, and controls the gas accretion and
  orbital migration of massive planets.
Recent observations and
  modeling show that the combination of spatially resolved
  imaging
and spectra at infrared {\it and} millimeter wavelengths is
  needed to make advances in this area. 
  
The eSMA will have the sensitivity, dynamic range and resolution to
enable the study of large and diverse samples of protoplanetary disks,
leading to key constraints on theoretical models of disk structure.
Three areas in which the eSMA will be particularly powerful include:
(i) benchmarking the nearest disks prior to ALMA; (ii) young disks in
the embedded phase of star formation, where the high angular
resolution and sensitivity is needed to separate disks and envelopes;
and (iii) transitional disks with large inner holes, perhaps
indicative of planet formation.  
For embedded sources, they will also shed light onto the launching
mechanism of the outflows and their relation to the properties of the
disks.

\item {\it High-mass star formation.} ---
The eSMA will help answer several
key questions in this area: (i) clustered star formation and the IMF;
(ii) circumstellar kinematics on $<$1000~AU scales;
(iii) pre-stellar phases.\\
The eSMA is very well suited to address these issues. Its high angular resolution is essential to
separate individual sources in distant (few kpc), crowded protostellar clusters, or
to map the structure and kinematics of the pre-stellar cores on
scales on which fragmentation is expected to occur. Its high sensitivity
is essential to measure a large range of masses and thus get a good sampling of the protostellar
mass distribution. Moreover, its high operating frequency makes for a better mass sensitivity per
unit flux density. 
Finally, in addition to providing mass determination from the continuum, eSMA observations
of line emission (such as CH$_3$OH, SiO, HCN, ions and deuterated species) 
could be used to measure the temperatures and densities of the sources, 
trace their molecular outflows, and measure their mid-infrared radiation field,
that is, to pinpoint physical and chemical condition in the cores.

\item {\it Evolved stars.} --- Studying the later stages of stellar
  evolution is important for an understanding of stellar populations
  and the chemical enrichment of galaxies, as well as the life cycle
  of stars and matter.  With its high spatial-resolving power at
  345~GHz, the eSMA will be an excellent facility to map the dust and
  kinematics in the envelopes of late-type stars, as well as
  investigate the magnetic field of their outflows.

\item {\it Nearby Galaxies: starbursts, mergers, and AGN.}  ---
Activity in galaxies is attributed to two main phenomena:
  extended starbursts and accretion
onto a central supermassive black
  hole to form an active galactic nucleus (AGN). Both types of
activity are fed by massive reservoirs of molecular gas and both
  may significantly affect their
surrounding clouds through their
  strong radiation fields. Studying the molecular gas in
  active galaxies requires both high
  angular resolution and high sensitivity, even
for very nearby AGN
  and starburst galaxies.  The eSMA will be of great help to tackle
  this task by bringing new information in several topics: 
(i) separating AGN and starburst activity in Ultra-luminous infrared
  galaxies (ULIRGs); (ii) physical and chemical conditions of dense
  molecular gas in the heart of ULIRGs; (iii) excitation conditions in
  nearby active galaxies; (iv) detailed views of the inner few hundred
  parsec of nearby galaxies.

An important aspect of this program
will be to combine new eSMA observations with existing SMA
data to trace the emission on a broad range of spatial scales.
Also, the high spatial resolution of the eSMA will make it
possible, for the first time, to match molecular line observations to
near-infrared ionic line observations in nearby galaxies.

\item {\it High-redshift galaxies.}--- 
Submillimeter
  studies of high-redshift galaxies have been a key area of activity
  for all three of
the observatories involved in the eSMA. Five topics have
  been identified where the eSMA can 
provide important new
  insights into the properties and nature of the important population
  of high-redshift submillimeter galaxies (SMGs):
(i) Exploring the far-IR/radio correlation on kpc scales at high redshift;
(ii) Beyond the confusion limit: imaging SMGs representative of the cosmic
far-IR background;
(iii) Imaging cold dust reservoirs in massive high-redshift AGN: witnessing the
birth of massive ellipticals;
(iv) Imaging bright SMGs with multiple radio counterparts: testing simulations
of galaxy-galaxy mergers;
(v) Imaging SMGs without radio counterparts: is there a significant population
of distant, massive starbursts?

\end{itemize}

\section{FIRST eSMA IMAGE OF A SCIENCE OBJECT: IRAS16293--2422}

IRAS16293--2422 (hereafter IRAS16293) is a well-known young, solar-mass
protostar, which consists, at arcsec resolution level,
 of two main components (IRAS16293A and IRAS16293B).
 IRAS16293 was observed on 2008 March 13 with 
 the SMA in the compact configuration. Despite the
 far from optimal eSMA beam, the eSMA
 team chose to proceed with the observations because
 the good weather conditions and sufficient
 brightness of IRAS16293 would allow the data to be self-calibrated,
 thereby obtaining the first science-verification data.
 
\begin{figure}[ht]
   \begin{center}
   \begin{tabular}{c}
   \includegraphics[width=0.9\textwidth]{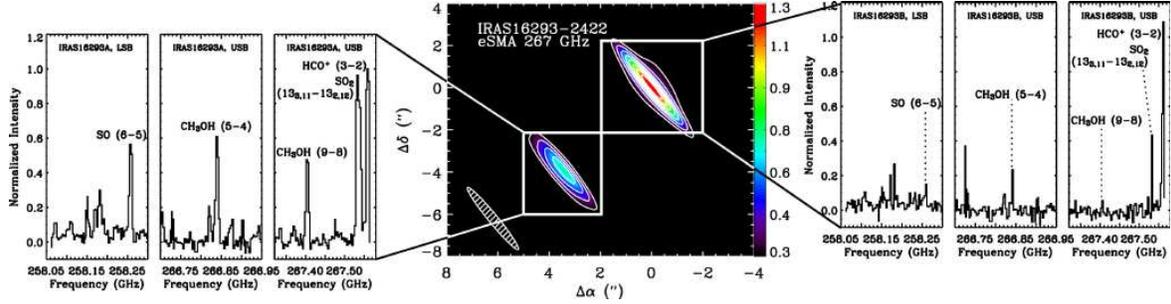}
   \end{tabular}
   \end{center}
   \caption[] 
  { \label{cont i16293} 
   eSMA continuum map and spectra of IRAS16293. 
   Absolute astrometry information is lost during the self-calibration process, 
so that the offsets here are relative to the phase center,
$\alpha$(2000) = $16^{\rm h}32^{\rm m}22\farcs 62$,  
$\delta$(2000) = $-22^{\circ}28'32\farcs 32$.
The 0\farcs38$\times$3\farcs54 beam is shown in the lower left corner. ---
The left and right panels show, for selected frequency ranges in the LSB and USB,
the spectra averaged over the box regions around the SE and NW components respectively.
Intensities are normalized to the 
strongest line and major transitions (or their positions) are labeled. 
 }
  \end{figure}

 Figure~\ref{cont i16293} shows the continuum as observed by the eSMA
 at 267~GHz. Although the absolute astrometry is lost in the process of self-calibration,
 the relative positions of the two components are consistent with the literature.
 Moreover, despite the facts that no high-spectral resolution data could be taken and that
 flux calibration could not be performed, the spectra
 displayed in Fig.~\ref{cont i16293} reproduce qualitatively well the
 particularities reported in the literature for this source, namely that the SE component
(the weaker component in the continuum in the 1~mm wavelength range) 
has broader and stronger lines than the NW component\cite{bottinelli-etal04-iras16293,kuan-etal04,bisschop-etal08} (Note that HCO$^+$ seems brighter in the NW component, 
but this is probably due to the fact that this line is located near the edge of a band).
 
\section{FIRST SCIENTIFIC RESULTS} 

\subsection{\pks: first direct absorption measurement 
of C/CO in a normal galaxy}
\label{pks}

\pks\ is a radio loud quasar with a redshift of $z$ = 2.507 
\cite{lidman-etal99}. 
Millimeter continuum emission images of \pks\ show two compact components, 
north-east (NE) and south-west (SW), separated by $\sim1''$. 
This double-peaked structure represents two images of the background quasar, 
magnified and distorted by a lensing system at $z$ = 0.88582  
\cite{wiklind+combes96}.  
This lens was indirectly detected by the observations of broad \hi\ and 
molecular absorptions in the millimeter spectra of \pks\
(e.g., Ref.~\citenum{wiklind+combes96,wiklind+combes98,gerin-etal97}).
Ref.~\citenum{wiklind+combes98} suggested it to be a spiral galaxy, which 
was recently confirmed by direct HST and IR images  
\cite{courbin-etal02,winn-etal02}. 
Observations of absorption lines reported in Ref.~\citenum{wiklind+combes98,muller-etal06}
show that the molecular absorption is most conspicuous towards the SW component, with
only a weaker absorption feature detected  in association with 
the NE component.

As shown by previous studies, the line of sight towards \pks\ offers a
unique opportunity to study the physics and chemistry of the gas in
a random molecular cloud in 
the lensing galaxy. Although several molecular species (CO, CS, HCN,
HNC, HCO$^+$, H$_2$S, H$_2$CO, N$_2$H$^+$, and some of their
isotopologues) have been detected in this source, little is know about
the atomic content.  However, this is important as atoms play a
dominant role in the energy balance of the gas, and observations of
atomic species such as \ci\ can be used to determine the physical and
chemical conditions of the gas.
In particular, the [\ci]/[CO] is often used as a diagnostic of the
type of PDRs from which they originate,
but such analyses refer to
emission line data averaged over large areas and suffer
from radiative transfer and excitation effects. The current eSMA data probe
the C/CO abundance ratio directly for the first time in absorption in
a dense molecular cloud.
Such a measurement has not yet been possible in our own Galaxy because of
the difficulty of doing subarcsec interferometry at 492~GHz. The C/CO
ratio is also an important ingredient for testing models for the formation
of more complex carbonaceous molecules.
\\

\subsubsection{eSMA observations}
Observations of \pks\ 
($\alpha$(2000) = $18^{\rm h}33^{\rm m}39\farcs 889$,  
$\delta$(2000) = $-21^{\circ}03'39\farcs 77$) 
were carried out with the eSMA 
on 2008 April 14 (with eight SMA antennas in the
very-extended configuration, vex), 
targeting the ($^3$P$_1-^3$P$_0$) transition of \ci\ at a rest frequency of 492.161 GHz.
Table~\ref{tab:obs} summarizes the observed frequencies, spectral resolution, 
and resulting beam sizes. See Ref.~\citenum{bottinelli-etal08-pks1830}
for a description of the data calibration and reduction process.

\begin{table}[h]
\centering
\begin{minipage}{0.9\textwidth}
\begin{center}
\label{tab:obs}
\caption{Observational parameters for \pks }
\begin{tabular}{|l|c|c|c|}
\hline
Parameter & \ci, eSMA & continuum, eSMA & continuum, SMA \\
\hline
Observed frequency (GHz)$^a$ & 260.97965 & 267 & 267 \\
\hline
Channel width &  0.93 km s$^{-1}$ & 2 $\times$ 1.5 GHz$^b$ & 2 $\times$ 2 GHz\\
\hline
Beam size (FWHM) and P.A. & $0\farcs62 \times 0\farcs30$, $38^\circ$
& $0\farcs55 \times 0\farcs22$, $32^\circ$ 
& $0\farcs50 \times 0\farcs33$, $25^\circ$\\
\hline
\end{tabular}
\end{center}

$^a$ {~Corresponding to a rest frequency of 492.16065~GHz, from 
$\nu_{\rm obs} = \nu_{\rm rest}/(1+z)$, with $z$ = 0.88582.} \\
$^b${~Note that in the 230~GHz band, the eSMA bandwidth is
limited by that of the JCMT's receiver A3.}
\end{minipage}
\end{table}

\subsubsection{Results}

\begin{figure}[ht]
\begin{center}
\vspace*{0.1cm}
   \includegraphics[width=0.9\textwidth]{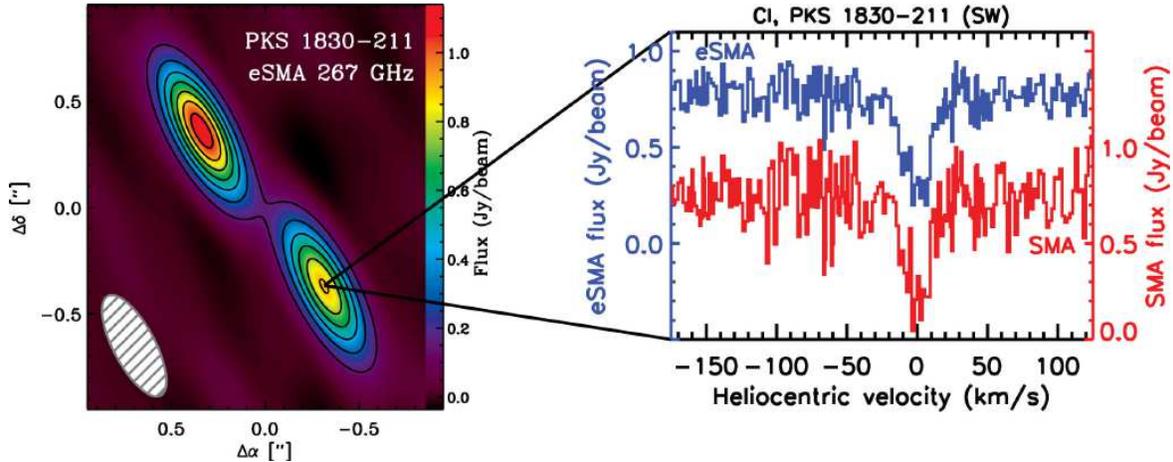}
 \end{center}
\caption[]{ \label{cont} 
eSMA continuum map and spectra of \pks.
Absolute astrometry information is lost during the self-calibration process, 
so that the offsets here are relative to the center of the double point source.
The contour levels are multiples of 150 mJy. 
The 0\farcs55$\times$0\farcs22 beam is shown in the lower left corner.
The spectral windows show the eSMA (top, flux scale on the left y-axis) and 
SMA (bottom, flux scale on the right y-axis)
spectra of \ci\ ($^3$P$_1-^3$P$_0$)
toward the SW component.
Note the improvement in noise level between the eSMA (rms = 0.085~\jyb) 
and SMA (rms = 0.143~\jyb) data.  
   }
\end{figure}

The left panel of Fig. ~\ref{cont} shows the continuum map obtained with the eSMA, derived 
from line-free channels.  
\ci\ absorption is only seen toward the
southern component as shown in the right panel of Fig.~\ref{cont}.
This illustrates the importance of sub-arcsecond data: the optical depth
determination could be off by more than a factor of 2 if the two components
were unresolved.
Moreover, the comparison of the profiles obtained with the eSMA and SMA-only
displayed in the right panel of Fig.~\ref{cont} clearly shows the gain in 
sensitivity provided by the eSMA.

Together with additional SMA observations of CO absorption,
we therefore obtained the first absorption measurement of the
C/CO abundance ratio for an average line of sight in a dense molecular
cloud in a normal spiral galaxy. 
A preliminary analysis indicates that the line profiles can be fitted with
2 or 3 components, each of which have a C/CO ratio of $\sim$1.
This value indicates that the region of the lensing spiral galaxy at $z$=0.889
intervening in the line of sight is mostly translucent cloud material.

\subsection{IRC+10216}

\begin{figure}[h]
\begin{center}
 \begin{tabular}{cc}
\includegraphics[height=4.5cm]{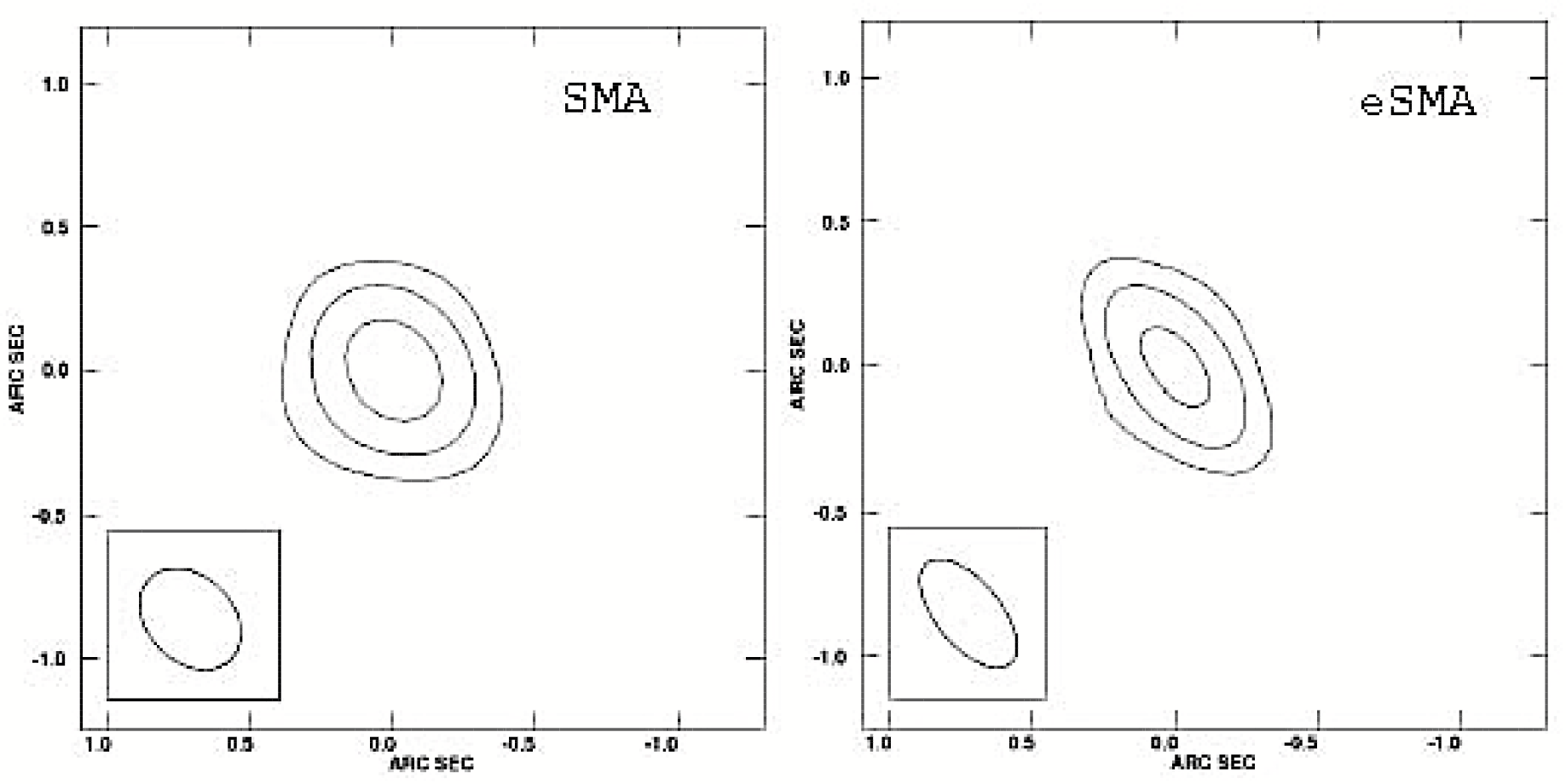} 
& 
\includegraphics[height=4.5cm]{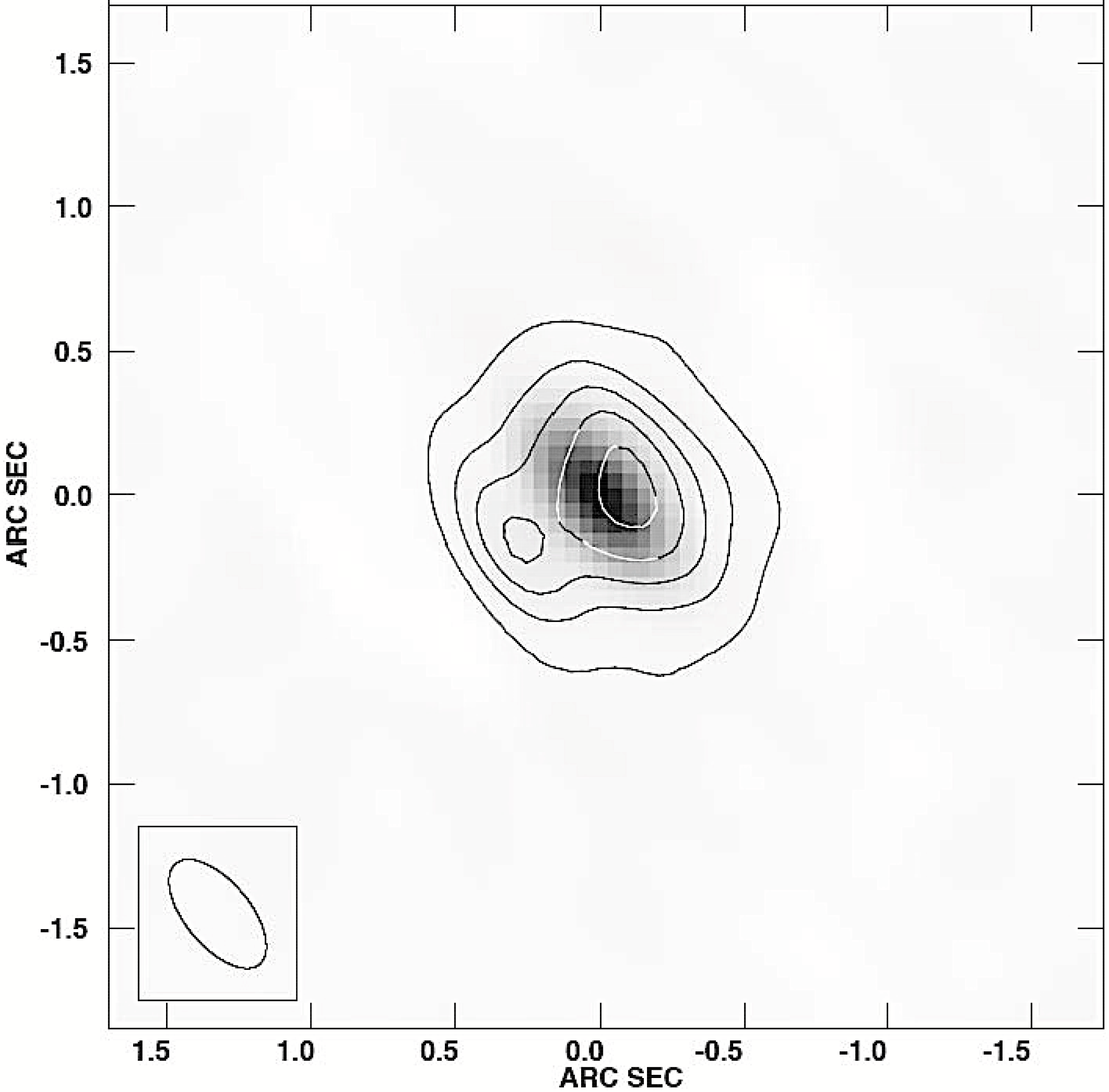} 
 \end{tabular}
   \end{center}
  \caption[]{ \label{fig:irc10216} {\it Left, middle:} SMA and eSMA continuum images of IRC+10216. 
  The contours are drawn at 3, 9, and 27 $\sigma$.
Beam sizes are shown in the lower left corner and are
$0\farcs40\times0\farcs30$ and $0\farcs46\times0\farcs22$ for the SMA and eSMA 
  respectively\cite{shinnaga-etal08-irc10216}. ---
  {\it Right:} eSMA map of the HCN $J=3-2$, $v$=(0,1,0) transition (contours) overlaid on
  top of the continuum emission of the central star (gray scale).
  The line emission was averaged over 
all channels for that transition.  
The contours are drawn at 3,
9, 15, 21, and 27 $\sigma$\cite{shinnaga-etal08-irc10216}.  
  }
   \end{figure}

IRC+10216 is the best-studied carbon star that shows a very rich spectrum of 
molecular lines at millimeter and submillimeter wavelengths. 
Hydrogen cyanide (HCN) is known to show maser components towards a few 
carbon-rich evolved stars including IRC+10216.  
Previous observations at the CSO revealed that the HCN higher $J$ transition ($J=9-8$) 
of IRC+10216 even shows laser  action \cite{schilke-etal00}.   
eSMA observations of IRC+10216 were performed on 2008 April 14,
with the correlator tuned to target the HCN $J=3-2$ components 
in the $v$=(0,1,0) state\cite{shinnaga-etal08-irc10216}. 
The left and middle panels of Fig.~\ref{fig:irc10216} show a comparison
of the continuum images taken with the SMA and with the eSMA, respectively.
Signal-to-noise is improved by about 67\% when we use the eSMA.  The  observations allowed us to spatially resolve the maser clump features for the first time as shown in the right panel of
Fig.~\ref{fig:irc10216}.  

\section{CONCLUSION}

In the past two years, substantial progress has been achieved to make
the eSMA into a fully functional instrument. Thanks to the dedication
and commitment of the commissioning team and of all three
observatories, the eSMA is now moving in the fast lane towards
harvesting exciting new results, as demonstrated by the first
scientific output of this array.  Moreover, the numerous scientific
areas to which the eSMA can make a significant contribution provide a
bright perspective for the coming years.

\appendix    

\section{Phase dependence on baseline components and antenna-dependent terms}

\subsection{Baseline components (positions)}
\label{bl eq reg}

\noindent The phase of antenna $i$, w.r.t. the reference antenna is given by:

\begin{equation}
\label{eq reg}
\phi_{i} = \frac{2\pi}{\lambda} 
 \bigg [ \Delta X_i \cos \delta \cos H 
- \Delta Y_i \cos \delta \sin H 
+ \Delta Z_i \sin \delta 
 \bigg ]
\end{equation}

where 
$\delta, H$, and $\epsilon$ = declination, hour angle and elevation of the source,
$\Delta X, \Delta Y, \Delta Z$ are the uncertainties in the $X, Y, Z$ positions of antenna $i$.

\subsection{Antenna-dependent terms: elevation}
\label{bl eq el}

Elevation terms can be added to equation \ref{eq reg}, assuming they are of the generic form 
$\Delta a \cos(\epsilon) + \Delta b \sin(\epsilon)$, we have:

\begin{equation}
\phi_{i} = \frac{2\pi}{\lambda} 
 \bigg [ \Delta X_i \cos \delta \cos H 
- \Delta Y_i \cos \delta \sin H 
+ \Delta Z_i \sin \delta 
+ \Delta a \cos \epsilon
+ \Delta b \sin \epsilon  \bigg ]
\end{equation}

\begin{figure}[h]
 \centering
\includegraphics[bb=0 0 529 200,width=0.5\textwidth]{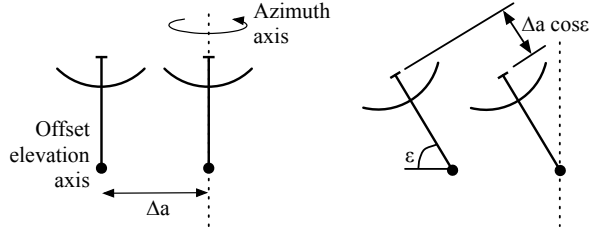}
\caption{\label{non-intersection} 
Schematic representation of the non-intersection of axes\cite{lay95}.}
 \end{figure}

\noindent We expect that the phase variations proportional to $\cos(\epsilon)$ are  
dominated by the non-intersection of the axes (see Fig.~\ref{non-intersection}), we identify $\Delta a$  
with the distance between the azimuth and elevation axes of antenna $i$, absorbing  
any other (smaller) effects with a $\cos(\epsilon)$ dependency into this term.

\bigskip

\noindent Now, if we look at the phase difference for two quasars $q_1$ and $q_2$:
\begin{eqnarray}
\phi_{i}(q_1) - \phi_{i}(q_2) = 
\frac{2\pi}{\lambda}  \bigg [  & ( \Delta X_i \cos \delta_1 \cos H_1 
- \Delta Y_i \cos \delta_1 \sin H_1 
+ \Delta Z_i \sin \delta_1 
+ \Delta a \cos \epsilon_1 + \Delta b \sin \epsilon_1 ) & \nonumber \\
- &
( \Delta X_i \cos \delta_2 \cos H_2 
- \Delta Y_i \cos \delta_2 \sin H_2 
+ \Delta Z_i \sin \delta_2 
+ \Delta a \cos \epsilon_2  + \Delta b \sin \epsilon_2 )  \bigg ] \label{phase diff}
\end{eqnarray}

 \noindent For Eq.~\ref{phase diff} to be independent of the $x$ and $y$ terms, 
a pair of two quasars are observed, such that $H_2=H_1$ and $\delta_2=-\delta_1$, 
which yields
$\cos \delta_2=\cos \delta_1, \cos H_2 = \cos H_1, \sin H_2 = \sin H_1$ and 
$\sin \delta_2 = -\sin \delta_1$.
We are left with the $\sin \delta$ and $\epsilon$ terms and
 Eq. (\ref{phase diff}) yields:
\begin{equation}
 \phi_{i}(q_1) - \phi_{i}(q_2) =\frac{2\pi}{\lambda} \bigg [ 2 \Delta Z_i \sin \delta_1
 + \Delta a(\cos \epsilon_1 - \cos \epsilon_2)  + \Delta b(\sin \epsilon_1 - \sin \epsilon_2)  \bigg ]
\end{equation} 

\section{Sensitivity ratios for dual vs single polarization with the eSMA}
\label{ap:sensitivity}

In this section, we assume that the SMA, JCMT and CSO antennas are the same, except for
their diameter (8, 15 and 10.4\,m respectively) 
and their aperture efficiency, $\eta_{\rm a}$ (0.67, 0.55 and 0.60 respectively).

For a homogeneous array, assuming a fixed integration time and bandwidth,
the noise level $\sigma$ is such that:
\begin{equation}
\sigma^2  \propto  \frac{T_{\rm sys}^2}{A_e^2~N_b~n_p} \\ 
\end{equation}
where $T_{\rm sys}$ is the system temperature,
$A_e=\eta_aA~=~\eta_a~\pi\,D^2/4$ is the effective aperture,
$\eta_{\rm a}$ the aperture efficiency, $A$ the geometric collecting area, 
$D$ the diameter of the antenna, 
$N_b=N_a(N_a-1)/2$ is the number of baselines, $N_a$ the number of antennas,
and $n_p$ = 1 or 2 for single or dual polarization respectively.\\

For a heterogeneous array, the denominator changes to the geometric mean over
the antennas:
\begin{eqnarray}
\displaystyle \sigma^2 & \propto 
& \frac{T_{{\rm sys},i}~T_{{\rm sys},j}}{n_p~\sum_{i=1,j=i+1}^{N(N-1)/2-1,N(N-1)/2}A_{e,i}A_{e,j}} \\
& \propto & \frac{1}{n_p}~\frac{T_{{\rm sys},i}~T_{{\rm sys},j}}{ N_{\rm S-S}(A_{e,{\rm S}}~A_{e,{\rm S}})
			     + N_{\rm S-J}(A_{e,{\rm S}}~A_{e,{\rm J}})
			     + N_{\rm S-C}(A_{e,{\rm S}}~A_{e,{\rm C}})
			     + N_{\rm J-C}(A_{e,{\rm J}}~A_{e,{\rm C}})
			     },\label{eq:sigma}
\end{eqnarray}
where $N_{\rm S-S},N_{\rm S-J},N_{\rm S-C},N_{\rm J-C}$
are the number of SMA-SMA, SMA-JCMT, SMA-CSO and JCMT-CSO baselines, and 
$A_{e,{\rm S}}, A_{e,{\rm J}}, A_{e,{\rm C}}$ are the effective apertures of the SMA, JCMT
and CSO antennas and are 18.94, 97.19 and 50.97~m$^2$ respectively.\\

Assuming identical $T_{\rm sys}$ for all antennas as a 0$^{\rm th}$ order approximation,
and using the relevant values in Tab.~\ref{tab:config} for the number of baselines, 
we therefore obtain the following noise level ratios:\\
eSMA$_{\rm dp}$ / eSMA$_{\rm sp}$ = 0.83,
SMA$_{\rm dp}$ / SMA$_{\rm sp}$ = 0.71, and
eSMA$_{\rm dp}$ / SMA$_{\rm sp}$ = 0.43,\\
corresponding approximately to an improvement in observing speed of a factor of $\sim$1.5, 2, and
5, respectively.

\begin{table}[h]
\centering
\begin{minipage}{0.85\textwidth}
\begin{center}
\caption{Number of baselines for Eq.~\ref{eq:sigma}}
\label{tab:config}
\begin{tabular}{|l|c|c|c|c|c|}
\hline
\multirow{2}{*}{\backslashbox{Config.$^a$}{Baseline type}}
&   SMA-SMA & SMA-JCMT & SMA-CSO & JCMT-CSO & Total \\
& & & & & \\
\hline
eSMA$_{\rm sp}$ & 8$\times$7/2 = 28 & 8 & 8 & 1  & 45\\
\hline
eSMA$_{\rm dp}$ & 6$\times$5/2 = 15 & 6 & 6 & 1  & 28 \\
\hline
\phantom{e}SMA$_{\rm sp}$ & 8$\times$7/2 = 28 & 0 & 0 & 0 & 28  \\
\hline
\phantom{e}SMA$_{\rm dp}$ & 8$\times$7/2 = 28 & 0 & 0 & 0  & 28 \\
\hline
\end{tabular}
\end{center}
$^a$ Configuration: note that due to correlator limitations, only six SMA antennas
can be used for the eSMA dual polarization mode.\\
\end{minipage}
\end{table}


\acknowledgments     

\noindent A number of people made the eSMA technically possible.  We
would like to thank all of them for their involvement and dedication,
in particular (but not limited to) Mark Bentum, Ken Brown, Jane
Buckle, Todd Hunter, Leo de Jong, Derek Kubo, Rob Millenaar, Alison
Peck, Glen Petitpas, Anthony Schinkel.  We also thank all the Leiden workshop
participants for stimulating discussions and for sharpening the eSMA
science case, in particular the panel leaders Darek Lis, Jonathan
Williams, Floris van der Tak, Christine Wilson, Wouter Vlemmings and
Rob Ivison. We are grateful to Ronald Stark for continued
support and encouragement. And last but not least, we are very much
indebted to Ray Blundell, Gary Davis and Tom Phillips, 
the directors of the SMA, JCMT and CSO, without whom the
eSMA would not be.

\noindent The development of the eSMA has been facilitated by a grant
from the Netherlands Organization for Scientific Research, NWO  
and NSF grant AST-0540882 to the Caltech  
Submillimeter Observatory.


\bibliography{bib_sb_spie}   
\bibliographystyle{spiebib}   

\end{document}